\documentclass[aps,preprint,amsmath,amssymb]{revtex4-1}
\usepackage{axodraw}
\usepackage{graphicx}
\newcommand{\nn}{\nonumber}
\newcommand{\bd}{\begin{document}}
\newcommand{\ed}{\end{document}}
\newcommand{\bc}{\begin{center}}
\newcommand{\ec}{\end{center}}
\newcommand{\be}{\begin{eqnarray}}
\newcommand{\ee}{\end{eqnarray}}
\newcommand{\ba}{\begin{array}}
\newcommand{\ea}{\ed{array}}
\newcommand{\strich}[1]{#1  \! \! \slash}
\newcommand{\eqn}{\global\def\theequation}
\newcommand{\sw}{sin^2 \theta_W}
\newcommand{\fbd}{f_B}
\renewcommand{\thefootnote}{\alph{footnote}}
\newcommand{\se}{\section}
\newcommand{\sse}{\subsection}
\newcommand{\bi}{\bibitem}
\def\figcap{\section*{Figure Captions\markboth
     {FIGURECAPTIONS}{FIGURECAPTIONS}}\list
     {Figure \arabic{enumi}:\hfill}{\settowidth\labelwidth{Figure 999:}
     \leftmargin\labelwidth
     \advance\leftmargin\labelsep\usecounter{enumi}}}
\let\endfigcap\endlist \relax
\def\reflist{\section*{References\markboth
     {REFLIST}{REFLIST}}\list
     {[\arabic{enumi}]\hfill}{\settowidth\labelwidth{[999]}
     \leftmargin\labelwidth
     \advance\leftmargin\labelsep\usecounter{enumi}}}
\let\endreflist\endlist \relax

\def\Journal#1#2#3#4{{#1} {{\bf #2},} {#4} {(#3)}}
\def\NCA{Nuovo Cimento}
\def\NIM{Nucl. Instrum. Methods}
\def\NIMA{{Nucl. Instrum. Methods} A}
\def\NP{{Nucl. Phys.} }
\def\NPB{{Nucl. Phys.} B }
\def\NPA{{Nucl. Phys. A}}
\def\PLB{{Phys. Lett.}  B}
\def\PL{{Phys. Lett.}}
\def\PPSA{{Proc. Phys. Soc.} A}
\def\PRP{{ Phys. Rep.}}
\def\PRL{ Phys. Rev. Lett.}
\def\PR{{Phys. Rev.}}
\def\PRD{{Phys. Rev.} D}
\def\PRC{{Phys. Rev.} C}
\def\ZP{{Z. Phys.}}
\def\ZPC{{Z. Phys. C}}
\def\EPJ{{Eur. Phys. J.}}
\def\EPJC{{Eur. Phys. J.} C}
\def\ZPA{{Z. Phys.} A}
\def\MPL{{Mod. Phys. Lett.}}
\def\MPLA{{Mod. Phys. Lett.} A}
\def\CPC{Comput. Phys. Commun.}
\def\JHEP{{J. High Energy Phys.}}
\def\JPG{{J. Phys. G.}}
\def\SJNP{Sov. J. Nucl. Phys.}
\def\NCA{ Nuovo Cimento}
\def\NIM{ Nucl. Instrum. Methods}
\def\NIMA{{ Nucl. Instrum. Methods} A}
\def\NP{{ Nucl. Phys.}}
\def\ANP{{Adv. Nucl. Phys.}}
\def\CPC{{Comput. Phys. Commun.}}

\begin{document}
\title
{\Large {\bf $\pi^{0}\to \gamma^*\gamma$ transition form factor within Light Front Quark Model}
}

\author{Chong-Chung Lih$^{1,2}$\footnote{E-mail address: 
cclih@phys.nthu.edu.tw} and 
Chao-Qiang Geng$^{3,2}$\footnote{E-mail address: geng@phys.nthu.edu.tw} 
}
\affiliation{
$^1$Department of Optometry, Shu-Zen College of Medicine and Management,
Kaohsiung Hsien,Taiwan 452   \\
$^2$Physics Division, National Center for Theoretical Sciences, Hsinchu, Taiwan 300\\
$^3$Department of Physics, National Tsing Hua University, Hsinchu, Taiwan 300 
}

\date{\today}

\begin{abstract}

We study
the transition form factor of $\pi^{0} \to \gamma^* \gamma$ as a function of the momentum transfer $Q^2$
within the light-front quark model (LFQM). We compare our result with the experimental data by BaBar as well as other 
calculations based on the LFQM in the literature. 
We show that our predicted form factor fits well with the experimental data, particularly those at the large $Q^2$
region.

\end{abstract}


\maketitle %

\se{Introduction}

The BaBar collaboration~\cite{BaBar} has reported a new data of the 
$\pi^{0} \to \gamma^* \gamma$ transition form factor $F_{\pi\gamma}(Q^{2})$ for the high  momentum transfer 
$Q^{2}$ up to $40$ GeV$^2$. 
To describe the data with the $Q^{2}$ dependence, the form factor is fitted to satisfy the formula
\be
Q^2 |F_{\pi\gamma}(Q^2)| = A\bigg(\frac{Q^2}{10\,\mathrm{GeV}^2}\bigg)^{\beta}\,
\label{1}
\ee
with $A=0.182\pm 0.002$ GeV and $\beta=0.25\pm 0.02$. 
Before the new data, most theoretical models predicted that the form factor approaches  the 
QCD asymptotic limit~\cite{pqcd}, depending on the pion distribution amplitude (DA)
with the $Q^{2}$ dependence under 10 GeV$^2$~\cite{asy,cz,bms}. 
Obviously, the experimental values for $Q^2>10$ GeV$^2$ by BaBar
are surprisingly much higher than the QCD asymptotic expectations 
and  thus, cannot be explained by the lowest perturbative results~\cite{pqcd}.
Even the high order corrections are considered~\cite{HOc1,HOc2}, the large $Q^2$ behavior is still hard to be understood. 
Recently, many proposals~\cite{BDA1,BDA2,BDA3,BDA4,BDA5,p0,p1,p2,p3,p4,p4a,p5,p6,p7,p8,p9,p10,p11,p12,p13,chpta,ldqcd,holQCD,p14} 
have been given in the literature to understand the  transition form factor, particularly the BaBar data for $Q^2 >10$ GeV$^2$.

In this note, we will use the phenomenological light front (LF) pion wave function 
to evaluate $Q^2 |F_{\pi\gamma}(Q^2)|$ in the light front quark model (LFQM)~\cite{vex1,lf1,lf2,lf3,lf4}. 
We will concentrate on the space-like region for the transition form factor. 
The LF wave function is manifestly boost invariant as it is expressed
in terms of the longitudinal momentum fraction and relative 
transverse momentum variables. The parameter in the hadronic 
wave function is determined from other information 
and the meson state of the definite spins can be constructed 
by the Melosh transformation. We emphasize that our 
derivation of the form factor can be applied to all allowed kinematic region. 
In Ref.~\cite{Hwang}, the study on the transition pion form factor based on
the LFQM has been done but the calculation for $Q^2$ is only up to 8 GeV$^2$. With the same set of parameters
in Ref.~\cite{Hwang}, the high $Q^2$ BaBar data cannot be fitted.
The use of the LFQM to understand the BaBar data 
has been explored in Ref.~\cite{LF-MF}. However, the conclusion in
Ref.~\cite{LF-MF} has failed to explain the data. In this work, we would like to revisit the LFQM to see if
it is indeed the case. 

This paper is organized as follows. In Sec.~II, 
we present the relevant formulas for the matrix element and form factor
for the $\pi^0 \to \gamma^* \gamma$ transition. 
In Sec.~III, we show our numerical analysis.
We give our conclusions in Sec.~IV.

\se{The form factor}

The transition form factor of $F_{\pi^{0}\to\gamma^*\gamma^*}(q^2_1,q^2_2)$, which describes 
the vertex of $\pi^{0} \gamma^*\gamma^*$, is defined by:
\be A(\pi^{0}(P)\to
\gamma^*(q_1,\epsilon_1)~\gamma^*(q_2,\epsilon_2))
=ie^{2}F_{\pi^{0}\to\gamma^*\gamma^*}(q^2_1,q^2_2)~\varepsilon_{\mu\nu\rho\sigma}~\epsilon^\mu_1
 ~\epsilon^\nu_2 ~q^\rho_1 ~q^\sigma_2\,, \label{def}
\ee
where $F_{\pi^{0}\to\gamma^*\gamma^*}(q^2_1,q^2_2)$ 
is a symmetric function under the interchange of $q^2_1$ and $q^2_2$. 
From the quark-meson diagram depicted in Fig.~1, the amplitude in Eq. (\ref{def}) is found to be 
\begin{figure}[htbp]
\includegraphics*[width=2in,height=6in,angle=-90]{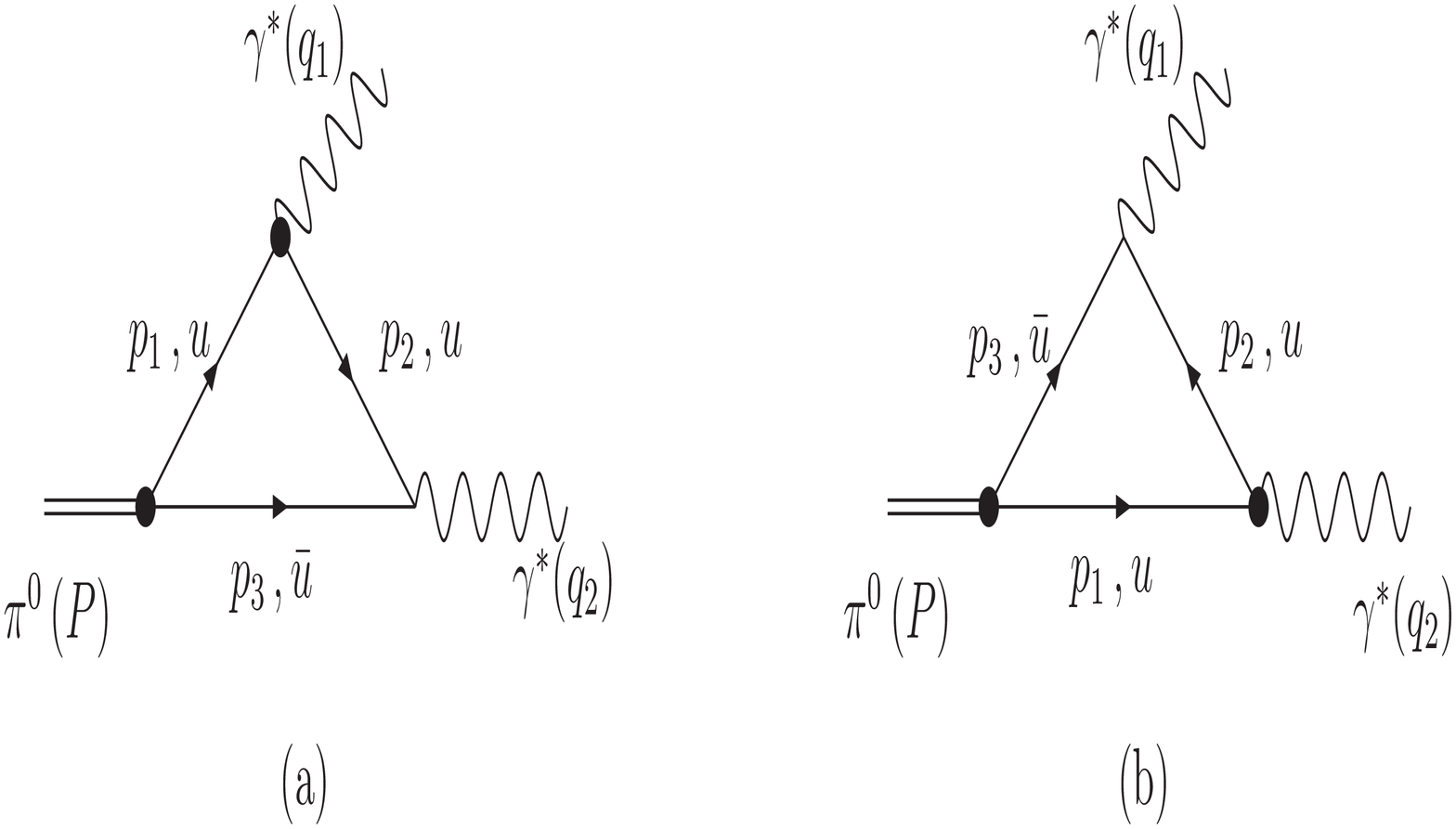}
\caption{  Loop diagrams that contribute to $\pi^{0} \to \gamma^*\gamma^*$.}
\label{F1}
\end{figure}
\be
A(Q\bar{Q}\to \gamma^*(q_1)~\gamma^*(q_2)) &=&
e_{Q}e_{\bar{Q}} N_{c}\int {d^4 p_3 \over{(2 \pi)^4}}
\Lambda_{P}\Bigg\{{\rm Tr}\Bigg[\gamma_5
        {i(-\not{\! p_3}+m_{\bar{Q}})\over{p_3^2-m^2_{\bar{Q}}+i\epsilon}}\not{\! \epsilon_2}
       {i(\not{\!p_2}+m_Q)\over{p_2^2-m^2_Q+i\epsilon}} \nn \\
&&\times \not{\! \epsilon_1}
        {i(\not{\! p_1}+m_Q)\over{p_1^2-m^2_Q+i\epsilon}}
\Bigg]+(\epsilon_1 \leftrightarrow \epsilon_2\,,\,q_1 \leftrightarrow q_2) \Bigg\}
 \nn \\
&&+(\,p_{1(3)} \leftrightarrow p_{3(1)}\,,\, m_{Q} \leftrightarrow
m_{\bar{Q}}) \,,
\label{matrix}
\ee
where $N_c$ is the number of colors, $e_{Q}$ is the quark electric charge 
and $\Lambda_{P}$ is the vertex function related to the 
$\pi^0$ meson bound state. To calculate the $\pi^0\to \gamma^*\gamma^*$ 
transition from factor within the LFQM, 
we have to decompose the $\pi^0$ meson into  $Q\bar{Q}$ Fock states, described as 
$(u\bar{u}-d\bar{d})/\sqrt{2}$. In the LF approach, the
LF meson wave function can be expressed by an anti-quark $\bar{Q}$ and a quark
$Q$ with the total momentum $P$  as:
\begin{eqnarray}
|M (P,S,S_z)\,\rangle&=& \sum_{\lambda _{1}\lambda_{2}}\int
[dp_{1}][dp_{2}]
2(2\pi)^{3}\delta ^{3}(P-p_{1}-p_{2})  \nonumber \\
&& ~~~~~~~~ \times \Phi_{M}^{SS_z}(z,k_{\bot}) b_{\bar{Q}}^{+}(p_{1},\lambda _{1}) d_{Q}^{+}(
p_{2},\lambda _{2}) |0\,\rangle\,,
\end{eqnarray}
where
\be
 [dp] = {dp^+d^{2}p_{\bot}\over 2(2\pi)^3 }\,,
\ee
 $\Phi_{M}^{\lambda _{1}\lambda _{2}}$ is the amplitude of
the corresponding $\bar{q}(q)$ and $p_{1(2)}$ is the on-mass shell
LF momentum of the internal quark.
In the momentum space, the wave function $\Phi_{M}^{SS_z}$ is given by
\be
\Phi_{M}^{SS_z}(k_1,k_2,\lambda_1,\lambda_2)
= R^{SS_z}_{\lambda_1\lambda_2}(z,k_\bot)~ \phi(z, k_\bot),
\label{phi1}
\ee
where $\phi(z,k_{\bot})$ represents the momentum distribution
amplitude of the constituents in the bound state and
$R^{SS_z}_{\lambda_1\lambda_2}$ constructs a spin state $(S,S_z)$
out of light front helicity eigenstates $(\lambda_1\lambda_2)$~\cite{melosh}.
The LF relative momentum
variables $(z,k_{\bot})$ are defined by
\be
&& p^+_1=z P^+, \quad p^+_2=(1-z) P^+\,,  \nonumber \\
&& p_{1\bot}=z P _\bot-k_\bot, \quad p_{2\bot}=(1-z)
P_\bot+k_\bot\,. \ee
The normalization condition of the meson
state is given by
\be
&&\langle M(P',S',S'_z)|M(P,S,S_z)\rangle =
2(2\pi)^{3} P^{+} \delta^{3}( P'- P)\delta_{S'S}\delta_{S'_zS_z}
\, , \ee
which leads the momentum distribution amplitude
$\phi(z,k_\bot)$ to
\be
N_c \int {dz\, d^2k_\bot\over 2(2\pi)^3}
|\phi(z,k_\bot)|^2 = 1\, .
\ee
We note that Eq.~(\ref{phi1}) can,
in fact, be expressed as a covariant form~\cite{vex1,vex2,lf1}
\be
\Phi_{M}^{SS_z}(z,k_{\bot })&=&\left( \frac{%
p_{1}^{+}p_{2}^{+}}{2[M_{0}^{2}-\left( m_{Q}-m_{\bar{Q}} \right) ^{2}]}\right)^{%
\frac{1}{2}}\overline{u}\left( p_{1}, \lambda _{1}\right)
\gamma^{5}v\left( p_{2},\lambda _{2}\right) \phi(z,k_{\bot}) \,,
 \nn \\
M_0^2&=&{ m_{\bar{Q}}^2+k_\bot^2\over z}+{ m_{Q}^2+k_\bot^2\over
1-z}\, .
\label{n6}
\ee
In principle, the momentum distribution
amplitude $\phi(z,k_\bot)$ can be obtained by solving the
LF QCD bound state equation~\cite{lf1}. However, before
such first-principle solutions are available, we would have to be
contented with phenomenological amplitudes. One example that has
been used is the Gaussian type wave function~\cite{lf2,lf3,lf4}:
\be
\phi(z,k_{\bot})=N\sqrt{\frac{1}{N_c}\frac{dk_{z}}{dz}} \exp
\left( -\frac{\vec{k}^{2}} {2\omega_{M}^{2}}\right) \,,
\label{7}
\ee
where $N = 4 ( \pi/\omega_{M}^{2})^\frac{3}{4}$, $\vec k =
(k_{\bot}, k_z)$, and $k_z$ is defined through
\be
z = {E_Q+k_z\over
E_Q + E_{\bar{Q}}} \,,~~ \ \ 1-z = {E_{\bar{Q}}-k_z \over E_Q + E_{\bar{Q}}} \, , ~~\ \
E_i = \sqrt{m_i^2 + \vec k^2} \,
\ee
by
\be
 & &
\ \ k_{z} =\left( z -\frac{1}{2}\right) M_{0}+\frac{m_{\bar{Q}}^{2}-m_{Q}^{2}}{%
2M_{0}}~\,,~~ M_0=E_Q + E_{\bar{Q}}\, .
\ee 
and $dk_z/ dz = E_Q E_{\bar{Q}}/ z(1-z) M_0$. 
After integrating over $p_3^-$ in Eq.~(\ref{matrix}), we
obtain 
\be 
A(Q\bar{Q}\to \gamma^*(q_1)~\gamma^*(q_2))
&=&e_{Q}e_{\bar{Q}} N_{c} \int^{q_{2}^{+}}_{0} dp_{3}^+ \int
{d^{2}p_{3\bot} \over 2(2\pi)^3\prod^3_{i=1} p^+_i} \bigg[
{\Lambda_{P} \over P^--p^-_{1{\rm on }}-p^-_{3{\rm on }}}
(I|_{p^-_3=p^-_{3{\rm on}}}) \nn \\
&& {1 \over q^-_2-p^-_{2{\rm on }}-p^-_{3{\rm on }}}
+(\epsilon_1 \leftrightarrow \epsilon_2,\,q_1 \leftrightarrow q_2) \bigg]
+(p_{1(3)} \leftrightarrow p_{3(1)}) \,,
\label{pole}
\ee
and
\be
I&=&{\rm Tr}[\gamma_5(-\not{\!p_3}+m_{\bar{Q}})\not{\! \epsilon_2}
(\not{\! p_2}+m_Q)\not{\! \epsilon_1}
(\not{\! p_1}+m_Q)]\,,~~~~~~p_{ion}^-={m_i^2+p_{i\bot}^2\over p_i^+}
\label{trace}
\ee
where the subscript $\{on\}$ stands for the on-shell particles.
One can extract the vertex function $\Lambda_{P}$ from
Eqs.~(\ref{matrix}), (\ref{n6}) and (\ref{pole}),
given by \cite{lf6,vex1,vex2}:
\be
\frac{\Lambda_{P}}{{P^--p^-_{1{\rm on }}-p^-_{3{\rm on }}}} &=&
{\sqrt{p_1^{+} p_3^{+}}\over \sqrt{2[M_{0}^{2}-\left( m_{Q}-m_{\bar{Q}} \right) ^{2}]}}\,\phi(z, k_\bot)~\,,
\ee
To calculate the trace $I$, we 
use the definitions of the LF momentum variables
$(z(x),k_{\bot}(k'_{\bot}))$ and take the frame with the transverse
momentum $(P-q_{2})_{\perp}=0$ for the $Q\bar{Q}$
state~($P$) and photon~($q_2$) in Fig.~1a. Hence, the relevant
quark variables are: 
\be
&&p_{1}^{+}=zP^{+},~~p_{3}^{+}=(1-z)P^{+},
~~p_{1\perp}=zP_{{\perp}}-k_\perp,~~p_{3\perp}=(1-z)P_{{\perp}}+k_\perp\,.
\nonumber \\
&&~p_{2}^{+}=xq_{2}^{+}, ~p_{3}^{+}=(1-x)q_{2}^{+},
~p_{2\perp}=xq_{2_{\perp}}-k^{'}_{\perp},
~p_{3\perp}=(1-x)q_{2_{\perp}}+k^{'}_{\perp}\,. \label{transmom}
\ee At the quark loop, it requires that \be
k_\perp=(z-x)q_{2_{\perp}}+k^{'}_{\perp}\,.
\label{momeq}
\ee
Take the trace $I$ into Eq.~(\ref{pole}) and consider the $\pi^0$ meson Fock states, 
the form factor $F_{\pi^0\to \gamma^*\gamma^*}(q^2_1,q^2_2)$ in Eq.~(\ref{def}) can be found to be: 
\be
F_{\pi^0\to \gamma^*\gamma^*}(q_{1}^{2},q_{2}^{2}) &=& -\frac{4}{3} \sqrt{N_{c}\over 6}
\int \frac{dx\,d^{2}k_{\bot }}{2\left( 2\pi \right) ^{3}}\bigg\{ {\Phi
        \left( z,k_{\bot }^{2}\right)}
\frac{m_{Q}+(1-z)m_{Q} k_{\bot}^{2}\Theta}{z(1-z)q_{2}^{2}-(m_{Q}^{2}+k_{\bot }^{2})}
        \nn \\
 &&~~~+(q_2 \leftrightarrow q_1)\bigg\}+(Q \leftrightarrow \bar{Q})  \,,
\label{fffv}
\ee
with
\be
\Phi (z,k_{\bot}^2) &=& N \sqrt{ {\frac{z(1-z) }{2 M_0^2}}}
\sqrt{{\frac{dk_{z}}{dz}}}\exp \left( -{\frac{\vec{k}^{2}}{%
2\omega_M^{2}}}\right)\,,  \nn \\
\vec{k}&=&(\vec{k}_{\bot}, \vec{k}_{z}) \,,  ~~
x=zr\,,~~  
\Theta = {\frac{1}{\Phi(z,k_{\bot}^2) }} {\frac{d\Phi(z,k_{\bot}^{2})}{%
dk_{\bot}^2}} \, ,  \nonumber \\
r&=&\frac{q_{2}^{+}}{P^+}=\frac
{(m_{\pi}^{2}+q_{2}^{2}-q_{1}^{2})+
\sqrt{(m_{\pi}^{2}+q_{2}^{2}-q_{1}^{2})^{2}-4q_{2}^2 m_{\pi}^{2}}}
{2m_{P}^{2}}\, \,.
\label{res}
\ee

\se{Numerical Result}

To numerically evaluate the transition form factor of $\pi^0\to \gamma^*\gamma$, 
we need to specify the parameters  in Eq.~(\ref{fffv}).
To constrain the quark masses of $m_{u,d,s}$ and the pion scale
parameter of $\omega_{\pi}$, we use the meson decay
constant $f_{\pi^0}$ and the decay branching ratio of $\pi^0 \to 2\gamma$,
given by the PDG~\cite{pdg}
\be
f_{\pi^0}&=&\,130\,{\rm
MeV},~~{\cal B}(\pi^0 \to 2\gamma)=\,(98.832\pm0.034)\,\%\,, \label{br2r}
\ee
 where the explicit expressions of $f_{\pi^0}$~\cite{fp} and ${\cal B}(\pi^0 \to 2\gamma)$ are
\be
f_{\pi^0}&=&\,4{\sqrt{N_c}\over\sqrt{2}}\int {dx\,d^2k_\perp\over 2(2\pi)^3}\,\phi(x,
k_\perp)\,{m\over\sqrt{m^2+k_\perp^2}}\,,
\ee
and 
\be
{\cal B}(\pi^0 \to 2\gamma)&=& \frac{(4\pi \alpha)^{2}
}{64\pi \Gamma_{\pi}} m_{\pi}^{3} |F(0,0)_{\pi^0\to 2\gamma}|^2 \,,
\ee
respectively.
As an illustration, we extracte $|F(0,0)_{\pi^0 \to 2\gamma}|=0.274$ in GeV$^{-1}$,
$m=m_u=m_d=0.24$ 
and $\omega_{\pi}=0.31$ in GeV, which will be used in our following
numerical calculations.

\begin{figure}[htbp]
\includegraphics*[width=4in]{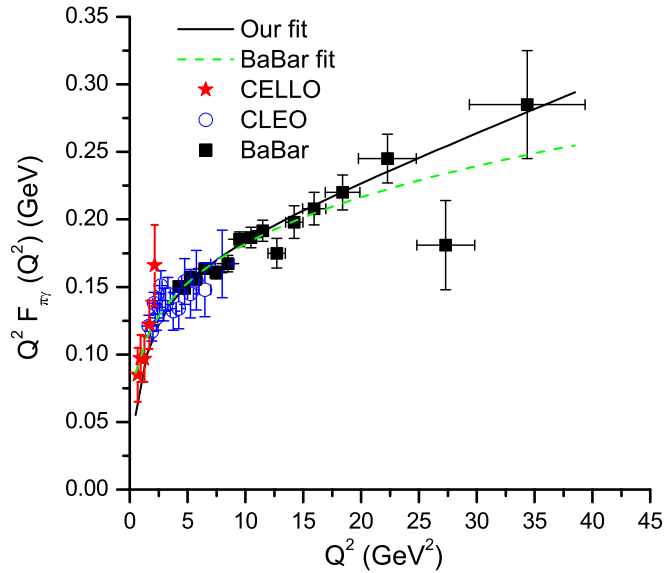}
\caption{ (Color online) $Q^2$ dependence of $F_{\pi\gamma}(Q^{2})$ in the LFQM. }
\label{F2}
\end{figure}

\begin{figure}[htbp]
\includegraphics*[width=4in]{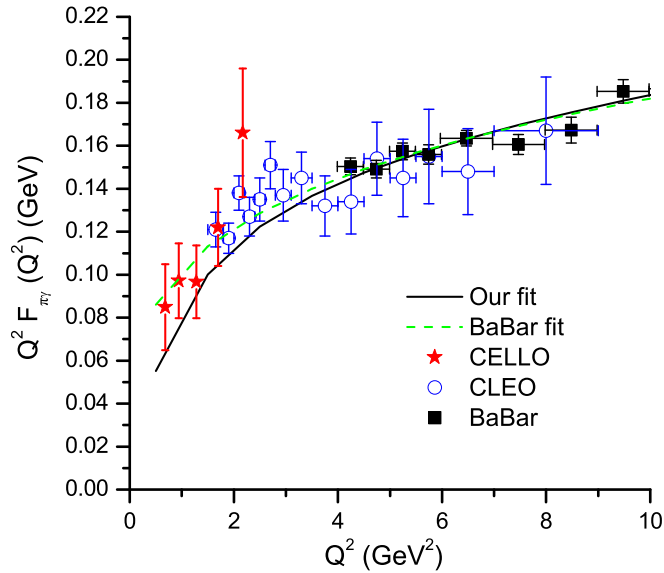}
\caption{ (Color online) $F_{\pi\gamma}(Q^{2})$ for $Q^2< 10$ GeV$^2$ in the LFQM. }
\label{F3}
\end{figure}

\begin{figure}[htbp]
\includegraphics*[width=4in]{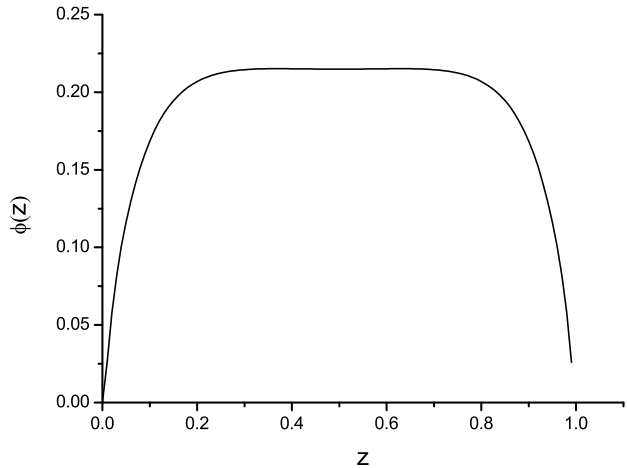}
\caption{ 
$\phi(z)$ as a function of $z$ in the LFQM. }
\label{DA}
\end{figure}

We now consider the case with  one of the photons on the mass shell. 
 From  Eq.~(\ref{fffv}),  the transition pion form factor becomes 
\be
F_{\pi\gamma}(Q^2)\equiv F_{\pi^0 \to \gamma^* \gamma}(Q^2,0)&=&\frac{4\sqrt{2}}{3}\sqrt{N_{c}\over 3}\bigg\{
        \int \frac{dx\,d^{2}k_{\bot }}{2\left( 2\pi \right) ^{3}} {\Phi
        \left( z,k_{\bot }^{2}\right)}
 \frac{m+(1-z)m k_{\bot}^{2}\Theta}{z(1-z)Q^{2}-(m^{2}+k_{\bot }^{2})}
\nn \\
&-&\int \frac{dx\,d^{2}k_{\bot }}{2\left( 2\pi \right) ^{3}} {\Phi
        \left( z,k_{\bot }^{2}\right)}
\frac{m+(1-z)m k_{\bot}^{2}\Theta}{(m^{2}+k_{\bot }^{2})}\bigg\}  \,.
\label{realff}
\ee
In Fig.~\ref{F2}, we  show the form factor in Eq.~(\ref{realff}).  
We note that the first term in Eq.~(\ref{realff}) dominates for the lower region of Q$^2$
and thus,  it can be use to describe the experimental data of BaBar~\cite{BaBar}, CLEO~\cite{cleo} 
and CELLO~\cite{cello} with Q$^2$ $\leq$ 10 GeV$^2$. 
The second one in Eq.~(\ref{realff}), related to the non-valence quark contributions,
is small for a small $Q^2$, but
it may enhance the form factor with a high value of Q$^2$.
As a result, we will include this term in our our numerical calculations.
To easily examine the Q$^2$ dependence of the form factor, we have 
fitted our result in terms of
the double-pole form:
\be
F_{\pi\gamma}(Q^{2})=\frac{F_{\pi\to \gamma\gamma}(0,0)}{M+(\beta Q)^{2}-(\alpha Q)^{4}}\,.
\ee
Explicitly,  we find that the dimension parameters of
 $\alpha=0.325$, $\beta=1.15$ and $F_{\pi^0\to \gamma\gamma}(0,0)=0.274$ in GeV$^{-1}$,
and the dimensionless parameter of $M=3.6$.
In Fig.~\ref{F3}, we concentrate  on the behavior of the form factor in the region with $Q^2< 10$ GeV$^2$. 
It is easy to see that our results fit the data well in this region similar to other theoretical calculations as expected. 
In Fig.~\ref{DA}, we show the DA, $\phi(z)$, as the function 
of the momentum fraction of the internal quark and meson longitudinal momenta, $z$, 
obtained by the integration of $k_{\bot}$ in Eq.~(\ref{7}).
%

As we mentioned in the introduction, the BaBar result cannot be fitted by extending the study in Ref.~\cite{Hwang} to a 
high $Q^2$. The main reason is due to the choices of the free parameters, such as the quark masses and $\omega_{\pi}$,
leading to a different sharp of the pion wave function.
Similarly, the main difference between our results and those in Ref.~\cite{LF-MF}  comes from DAs. In particular, 
our DA shown in Fig.~\ref{DA}
appears to be much broader.
We note that 
in Refs.~\cite{BDA1,BDA2,BDA3,BDA4,BDA5}, 
a more broader DA of the pion is utilized to fit the BaBar data,  particularly  in the high Q$^2$ region.
The results in these models differ slightly from ours  only at large values of Q$^2$. 
It is interesting to point out that our result  is 
almost identical with that in the Regge model~\cite{p7} and the double logarithmic behavior
 from the chiral anomaly effects~\cite{chpta}. 
Finally, we remark that the single data point at Q$^2=27.31$ GeV$^2$ by BaBar, which is apparently consistent with 
the QCD asymptotic limit, cannot be explained by this work within the framework of the LFQM.

\se{Conclusions}

We have studied the form factors of $\pi^0 \to \gamma^* \gamma$  within the LFQM. In our calculation, we have 
adopted the Gaussian-type wave function and evaluated the form factors for the momentum dependences 
in the all allowed Q$^2$ region. We have also parametrized the form factor in terms of the double-pole form. 
Our numerical values are  close to the experimental results by BaBar. 
In particular, our results  of the transition form factor fit well with the experimental data in the high Q$^2$ region,
which cannot be explained in the previous attempts based on the framework of the LFQM. 
Finally, we remark that due to the large  uncertainty  in the high Q$^2$ region for the BaBar data,
further theoretical studies as well as more precise experimental data are clearly needed. If some future experiment could not confirm the 
BaBar data but be rather in agreement with the QCD asymptotic limit, the parameters of 
the LFQM in this study should be either modified or ruled out.

\section{Acknowledgments}
This work was partially supported by National Center of Theoretical
Science and  National Science Council (NSC-97-2112-M-471-002-MY3 and NSC-98-2112-M-007-008-MY3) of R.O.C.

\end{document}